# Design Level Metrics to Measure the Complexity Across Versions of AO Software

Parthipan S[1], Senthil Velan S[2], Chitra Babu[3]
[1]PG Scholar
[1,2,3]Department of Computer Science and Engineering, SSN College of Engineering, Chennai, India
[1]parthipan1211@cse.ssn.edu.in, [2]senthilvelan@ssn.edu.in, [3]chitra@ssn.edu.in

*Abstract*—Software metric plays a vital role in quantitative assessment of any specific software development methodology and its impact on the maintenance of software. It can also be used to indicate the degree of interdependence among the components by providing valuable feedback about quality attributes such as maintainability, modifiability and understandability. The effort for software maintenance normally has a high correlation with the complexity of its design. Aspect Oriented Software Design is an emerging methodology that provides powerful new techniques to improve the modularity of software from its design. In this paper, evaluation model to capture the symptoms of complexity has been defined consisting of metrics, artifacts and elements of complexity. A tool to automatically capture these metrics across different versions of a case study application, University Automation System has been developed. The values obtained for the proposed metrics are used to infer on the complexity of Java and AspectJ implementations of the case study application. These measurements indicate that AspectJ implementations are less complex compared to the Java implementations and there by positively influencing the maintainability of software.

## I. INTRODUCTION

*IEEE* defines software complexity as the degree to which software has a design or implementation that is difficult to understand and verify [9]. During the evolution of a software, the client demands a better usable and highly modular software. If complexity is not addressed by a developer during the design stage of a software, then, the software could become even more complex during its evolution. Further, complexity is also an important factor to analyze the cost of developing and maintaining any software [6]. If the complexity of the software is kept low then it can have a positive impact on the understandability and modifiability during its evolution.

Aspect Oriented Software Development (AOSD) is a promising methodology that supports clear separation of core and cross-cutting concerns. A cross-cutting concern can affect many modules in a software. The common examples of cross-cutting concerns are functionalities encapsulating logging, security, persistence and exception handling. *AOSD* is used to achieve a higher modularity in software and modify the functionalities modeled as cross-cutting concerns without disturbing the core concern of the software. High quality *AO* software requires the designer to provide strong rationale behind the modularization of software during its evolution.

Measurement of software plays a major role in quantifying the characteristics of software. In order to measure complexity, the design property need to be measured using metrics capturing the symptoms that either increase or decrease the complexity in the artifacts of *AO* software. Hence, a set of software metrics need to be defined for the measurement. In this paper, metrics to capture the complexity arising out of the usage of artifacts in *AO* software have been proposed.

Software quality is a higher level attribute [2, 4, 5] that provides an understanding the internal strength and stability towards the use in time. Since, it is a higher level quality attribute, it is not possible to directly measure this but has to be inferred indirectly through the design properties of software. In this work, complexity has been considered design property and can be related to the higher level quality attribute, namely, maintainability.

In this paper, a new set of metrics has been proposed for the measurement of complexity in an *AO* software. In order to measure the values of the metrics, a tool named as Aspect Oriented Software Complexity Evaluation Tool *(AOSCE)* has been developed to automatically compute the complexity for the case study application. During automated calculation, the aspect and class constructs which contribute to the complexity are identified and evaluated. The rest of the paper is organized as follows. The motivation behind this work is explained in Section II. Existing work on complexity is discussed in Section III. The proposed complexity model is discussed in Section IV. The architecture of *AOSCE* tool with its internal modules are explained in Section V. A short description of the case study application is given in Section VI. The results of applying proposed metrics is explained in Section VII. The effect of complexity on maintainability for the case study application using the measured values of the metrics is discussed in Section VIII. Section IX concludes with scope for future enhancements.

## II. MOTIVATION

According to Bansiya [2], the number of methods within a class is directly proportional to the complexity of the class in a *OO* software. This is because if a class models more number of functions in the form of methods, there is a need for the designer to be aware of all these functions during evolution. Extending the measurement of complexity, Pataki [8] only considered the modeling of pointcuts and advices as indicators for complexity in *AO* software. However, there are other constructs in the *AO* artifacts that can positively or negatively influence complexity. Hence, in order to obtain a comprehensive understanding towards aspectization on complexity and intern on maintainability of *OO* software, there is a need to define a complexity evaluation model.

## III. EXISTING WORK

Kiczales [7] implemented seventeen out of the twenty three *GoF* design patterns in both Java and AspectJ programming languages. Based on this work, it was found that AspectJ implementation is able to reduce the complexity of implementation.

Pataki [8] measured the complexity of AspectJ implementation of the same set of design patterns using multi-paradigm metric, and found that complexity has decreased in *AO* implementation of adapter, builder, observer, and state patterns. The focus of measurement was only on two constructs of aspects and also not on the classes. Hence, there is a need to measure all constructs of *AO* that can influence the complexity of a software. The complexity of an AOP program depends on the both *OO* core concerns and the *AO* cross-cutting concerns.

Bansiya [2] proposed a quality model for object oriented design by measuring the constructs within the classes. This work has not been extended to the measurement of complexity of *AO* software. Hence, an *AO* software complexity evaluation model is needed to study the effect of aspectization in *OO* software.

## IV. PROPOSED COMPLEXITY MODEL

In order to understand the impact of using a new methodology for software development, it is imperative to design a model that can capture the core design properties. Since, the complexity property cannot be directly measured, a set of metrics is required to capture the design of constructs that either directly or indirectly affect the complexity. The proposed *AO* software complexity evaluation model is shown in Fig. 1.

### A. Artifacts

Artifacts are tangible products which are developed during the development of software. Based on this definition, in *AO*, the artifacts are pointcuts, advices, join points, classes and methods designed during development. Pointcut is a unique construct which can be used to capture a set of join points in the base or aspect code. Advice is the code segment that gets executed when the control reaches a join point, identified by a pointcut. A join point is a specific location in the execution of a program at which the advice can be woven into the code. A class is a container which contains both data members and its member functions. The role of these artifacts affecting the complexity property is explained in Section IV B.

### B. Proposed Metrics

Metrics are important in software engineering because it can be used to indirectly infer on high level quality attributes such as understandability, modifiability, complexity and maintainability [6]. Pataki [8] measured *AO* Complexity by using *AV* graph; the *AV* graph is based on the control flow and does not consider the data flow. Hence, the focus was only on the program complexity and was not on the data complexity [8]. The complexity was measured by considering only the pointcuts and advices while join points are not taken into account for calculating the software complexity. The available *AO* metrics consider only two constructs of aspects and are not suitable for the measuring constructs of classes. Chidamber and Kemerer (CK) [3] metrics were defined to measure the design complexity of *OO* software and to infer on higher quality attributes such as understandability and reusability. This work indicates that both *DIT* and *NOC* can together contribute to the measurement of the complexity of software.

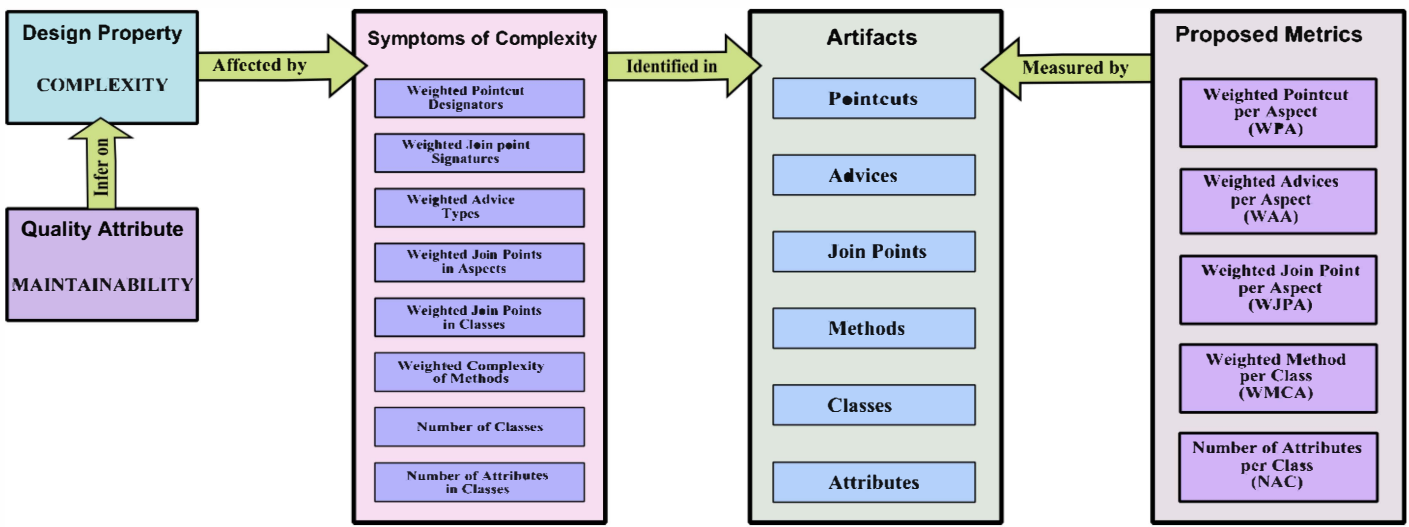

Fig.1. Aspect Oriented Software Complexity Evaluation Model

A quality model named as Quality Model for Object Oriented Design *(QMOOD)* was developed by Bansiya and Davis [2]. It is a model which evaluates a good number of internal quality attributes notably, Understandability and core design properties such as Complexity, Coupling, and Cohesion [2]. In this work, the complexity of an *OO* system is measured only by using one metric, namely, Number of Methods (*NOM*).

In *OO* software, the complexity of a class is calculated by adding the complexity of the methods and the complexity of attributes. In *OOP*, cross-cutting concerns are difficult to modularize, and hence the complexity of software is increased. Software metrics can be used to indicate the degree of system interdependence among modules and to provide valuable feedback for better understandability and reusability of software. Hence, a set of metrics are needed to measure the symptoms of complexity in *AO* software.

The complexity in *AO* software can be measured for its maintenance by means of calculating the value of the related metrics. The three artifacts, namely, pointcuts, advices and join points are used to measure the complexity of an Aspect.

*1) Weighted Pointcut per Aspect [WPA]:* WPA is calculated by adding the cognitive weight of the pointcut designator and cognitive weight of the join point signature used in an aspect. Pointcut designator describes when the advices are woven into the join points, e.g. `execution` designator signifies right before the execution of a method, and similarly `call` designator identifies the call to the execution of a method. In TABLE I, the commonly used designators are arranged in the order of the increasing degree of cognitive complexity. Cognitive complexity [1] stands for the complexity involved in understanding the weaving of an advice in a join point. Cognitive weights are assigned for each designator based on the cognitive complexity analysis and the weights are listed in TABLE I. The join point signature describes the functions that are related to the respective pointcut definitions and in our case, when `call()` and `execution()` designators are used in the pointcut. For example, `void||int *.func(..)` means that all the functions of all classes with name `func` irrespective of the number of arguments are the join points for the pointcut. The cognitive weight for join point signature is shown in TABLE II. The join point signature unqualified class name/method name has the highest weight as per the assignment of cognitive weights.

The formula to calculate *WPA(A)* is given in Equation 1.

$$WPA(A) = \sum_{i=1}^{m}[CW(PD_i) + CW(JS_i)] \qquad (1)$$

where,
*WPA(A)* is the Weighted Pointcuts per Aspect,
*CW(PD_i)* is the Cognitive Weight of Pointcut Designator for $i^{th}$ pointcut,
*CW(JS_i)* is the Cognitive Weight of Join point Signature for the $i^{th}$ pointcut, and
*m* is the number of pointcuts in Aspect *A*.

Considering *S* as a particular version of *AO* software taken for measurement, then WPA(S) for that version can be calculated using the formula given in Equation 2.

where,
*WAA(S)* is the *WAA* for a version of *AO* software, and

*3) Weighted Join Points per Aspect* [*WJP*]: A join point is a specific point at which advice can be woven into the code of

$$WPA(S) = \sum_{i=1}^{n}WPA(A_i) \qquad (2)$$

where,
*WPA(S)* is the *WPA* for a version of *AO* software, and
*n* is the number of Aspects in a version of *AO* software.

TABLE I. POINTCUT DESIGNATOR WITH COGNITIVE WEIGHT

| S. No. | Pointcut Designator (*PD*) | Cognitive Weight [*CW(JS)*] |
|---|---|---|
| 1 | execution | 0.1 |
| 2 | call | 0.2 |
| 3 | get | 0.3 |
| 4 | set | 0.4 |
| 5 | handler | 0.5 |

TABLE II. JOIN POINT SIGNATURE WITH COGNITIVE WEIGHT

| S. No. | Advice Type (AT) | Cognitive Weight [CW(AT)] |
|---|---|---|
| 1 | before() | 0.1 |
| 2 | after() | 0.1 |
| 3 | around() | 0.2 |

*2) Weighted Advices per Aspect [WAA]:* The value of *WAA* metric is calculated as the sum of cognitive weight of all advice types in an aspect. The advices are generally classified into three types, namely, `before()`, `after()` and `around()` advice. The cognitive weights assigned for the advice types are based on the cognitive complexity as shown in TABLE III. The `before()` and `after()` advices are less complex compared to the `around()` advice. Since `around()` advice can optionally bypass the execution of a join point, the cognitive weight assigned to it has a higher value compared to other two advice types.

TABLE II. ADVICE TYPES WITH COGNITIVE WEIGHT

| S. No. | Advice Type (AT) | Cognitive Weight [CW(AT)] |
|---|---|---|
| 1 | before() | 0.1 |
| 2 | after() | 0.1 |
| 3 | around() | 0.2 |

The formula to calculate *WAA(A)* is given in Equation 3.

$$WAA(A) = \sum_{i=1}^{m}CW(AT_i) \qquad (3)$$

where,
*WAA(A)* is the Weighted Advices per Aspect,
*CW(AT_i)* is the Cognitive weight for Advice Types in an Aspect, and
*m* is the number of advices in the Aspect A.
The formula to calculate *WAA(S)* is given in Equation 4.

$$WAA(S) = \sum_{i=1}^{n}WAA(A_i) \qquad (4)$$

*n* is the number of aspects.

an application. A join point is a well defined location in code that realizes both core and cross-cutting concerns. Through the type of constructs (`call, execution, etc.`) in the join point, it is possible to identify the set of qualified join points. For example, if the pointcut designator is `call` then method calls are qualified join points. Similarly, for the `execution` pointcut designator, the point right before the execution of the qualified method acts as a join point. In *AOP*, aspects can also have a set of join points related to other aspects. Hence, the proposed metric includes both classes and aspects during calculation. The *WJP* per class or aspect is the sum of cognitive weights of types of join points shadow in classes or aspects. The cognitive weight assigned to the identified designators based on its cognitive complexity is given in TABLE IV.

The formula to calculate *WJP(A)* is given in Equation 5.

$$WJP(A) = \sum_{i=1}^{m} CW(JA_i) + \sum_{j=1}^{n} CW(JC_j) \quad (5)$$

where,
*WJP(A)* is the Weighted Join Points per Aspect,
*CW(JA$_i$)* is the Cognitive Weight for the Join Points in Aspect,
*CW(JC$_j$)* is the Cognitive Weight for the Join points in Class,
*m* is the number of Aspects, and
*n* is the number of Classes.

TABLE IIII. JOIN POINT TYPES WITH COGNITIVE WEIGHT

| S. No. | Join point Type (JA/JC) | Cognitive Weight [CW(JA/JC)] |
|---|---|---|
| 1 | method execution | 0.1 |
| 2 | method call | 0.2 |
| 3 | exception handling | 0.3 |
| 4 | within advice | 0.4 |
| 5 | Attribute | 0.5 |
| 6 | particular method | 0.6 |
| 7 | particular class | 0.7 |
| 8 | particular package | 0.8 |
| 9 | control flow | 0.9 |
| 10 | boolean or combined | 1.0 |

*4) Weighted Methods per Class and Aspect [WMCA]*: *WMCA* is a metric that calculates the complexity of classes and aspects. In this metric, cognitive weight for each method is considered as equal. A cognitive weight of 1 is assigned for all methods in the class and aspect. If the number of methods in a class is high, then the coupling factor is increases because the modules are dependent and hence the class becomes complex. This metric is proposed by Chidamber and Kemerer and is used in calculating complexity for *OO* software.

The formula to calculate *WMCA(C)* is given in Equation 6.

$$WMCA(C) = \sum_{i=1}^{m} CW(MC_i) \quad (6)$$

The formula to calculate *WMCA(A)* is given in Equation 7.

$$WMCA(A) = \sum_{i=1}^{n} CW(MA_i) \quad (7)$$

where,
*CW(MC$_i$)* is the Cognitive Weight of Method Complexity in Class,
*CW(MA$_i$)* is the Cognitive Weight of Method Complexity in Aspect,
*WMCA(C)* is the Weighted Method per Class,
*WMCA(A)* is the Weighted Method per Aspect,
*m* is the Number of Classes, and
*n* is the Number of Aspects.

The formula to calculate *WMCA(S)* is given in Equation 8.

$$WMCA(S) = WMCA(C) + WMCA(A) \quad (8)$$

where,
*WMCA(S)* is the *WMCA* for a version of *AO* software.

*5) Number of Attributes per Class [NAC]*: *NAC* is one of class metric, to capture the spread of attributes in the classes of *OO* software. The metric can be defined as the ratio of the number of attributes in a software compared to its total number classes. This metric can be useful to infer on the following points.

- If a class is defined with more attributes, then it will exhibit a higher degree of coincidental cohesion and hence further decomposition is needed in order to reduce the complexity.
- If a class has no attributes, then design of other classes should be relooked, in order to verify whether the class with no attributes is unnecessarily accessing the attributes of other classes.

The formula to calculate *NAC* is given in Equation 9.

$$NAC = NA / NC \quad (9)$$

where,
*NAC* is the number of attributes per Class,
*NC* is the number of Classes, and
*NA* is the number of attributes in all Classes.

V. ARCHITECTURE OF AOSCE TOOL

The automation to apply the proposed metrics on the case study application requires a tool. This requirement is achieved by developing a Java based automated metrics measurement tool, namely, Aspect Oriented Software Complexity Evaluation *(AOSCE)* tool. The architecture of *AOSCE* tool is shown in Fig. 2. The tool consists of four modules namely Main module, File Identifier and Logger, Signature Extractor and Logger and Metric calculator and Logger. The functionalities of the modules are explained in the following sections.

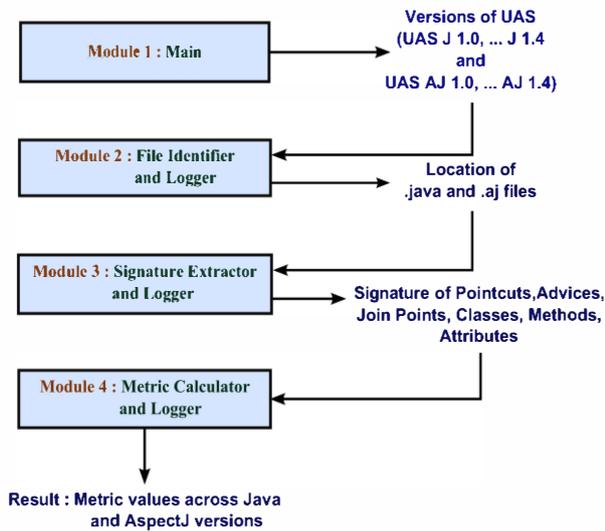

Fig. 2. Architecture of Proposed AOSCE Tool

### A. Main Module

The functionality of the module is to choose the directory of the files for the case study application. This directories contain the Java and AspectJ versions of *UAS*. The directory for each version contains both *.java* and *.aj* files implementing the functionalities in that version.

### B. File Identifier and Logger

The directory path is given as input to the File Identifier and Logger module. The directory is traversed to identify all the *.aj* and *.java* files. Once all the files are identified, a log file is created that contains the absolute path of both *.aj* and *.java* files. This log file will be used by the next module to extract the relevant details.

### C. Signature Extractor and Logger

The log file created in the previous module is given as input to the Signature Extractor and Logger module. Signature extractor will open and read all contents in each file identified and written in the log file by using the buffered reader. The contents of the opened file are traversed to check whether the keywords (pointcut, advice, void, class, aspect, etc.) are matched in the file. If a keyword is found then the signature is printed at the end of the same log file created in the previous module. Note that the keywords should be stored in this module before executing an application. The following elements are printed in the log file:

- Location of the .java file
- Signature of the classes
- Signature of the methods
- Signature of attributes
- Location of the .aj file
- Signature of the aspects
- Signature of the pointcuts
- Signature of the advices
- Signature of the join points

### D. Metric Calculator and Logger

The Metric Calculator and Logger module reads the contents of the log file created in the previous module. The metric values are calculated based on the elements identified in the log file. Once the metric values for the case study application that was given as input are calculated, the values are printed at the end of the log file. The formula to calculate the proposed metrics is clearly explained in Section IV B.

### VI. CASE STUDY – UNIVERSTIY AUTOMATION SYSTEM

An SOA application, with many core concerns and cross-cutting concerns, is required to understand the impact of complexity on maintainability of *AO* software. This requirement is satisfied by the University Automation System *(UAS)* Application that automates the operations of a typical university, since it possesses many scattered and tangled concerns. Hence, *UAS* is selected as the case study to understand the impact of complexity in *AO* software. Process flow diagram for the case study *UAS* is shown in the Fig. 3.

*UAS J 1.0* consists of only core concerns. *UAS AJ 1.1* consists of two cross-cutting concerns such as logging and persistence. Logging aspect is encapsulated to the Login web service of both the student and staff, in order to record who has logged-in and the timing when the user logs into the system. Persistence aspect is encapsulated to the Register web service of both the user, in order to store the provided details persistently in the database. In logging aspect, audit log and event log gets inherited from file log. Audit log captures the authenticated information. Event log logs the timing events of login and logout. In persistence aspect *SQL* translation gets inherited from persistence.

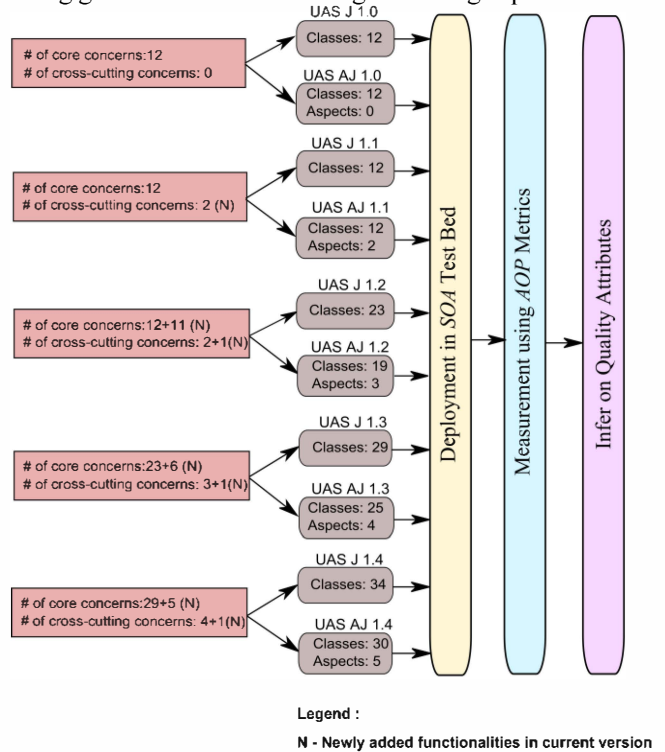

Fig. 3. Applying AO across versions of UAS

One more cross-cutting concern security is added to the version *UAS AJ* 1.2. Security is identified as the cross-cutting concern in this version. Because a single login concern is scattered for different users as different account. So this authentication functionality can be separated as security aspect of *AOP*. Login security and database security will inherit from the security aspect. User authentication has been performed in login security. Database authentication has been done in database security aspect.

*UAS AJ 1.3* consists of observer pattern as cross-cutting concern. Once the result has been updated, the student can be notified that the result has been updated. For this purpose, Observer pattern can be used.

*UAS AJ 1.4* consists of exception handling as cross-cutting concern. The implementation of Exception handling in traditional *OOP* results in tangling of exception handling concern and primary concern. This implicit coupling between concerns leads to maintenance and evolution problems. To avoid this, code to handle exceptions in *UAS* can be separated as aspect of *AOP* so that whenever exceptions are thrown in *UAS*, those exceptions can be effectively handled in this version of *UAS*.

## VII. MEASUREMENTS AND RESULTS

The Java and AspectJ versions of *UAS* are given as input to the *AOSCE* tool. The *AOSCE* tool calculates the values of the proposed metrics for both Java and AspectJ versions. The obtained values of the measurement are given in TABLES V and VI.

In the measurement the values of the proposed metrics, *WPA*, *WAA* and *WJP* are zero in all the Java versions of *UAS*. This is because of the metrics listed above only captures the constructs of aspects and there are no aspect constructs in the versions. During the evolution of software, new features are added to the version which is reflected by the increase in the value of *WMCA* over versions.

## VIII. DISCUSSION ON MEASUREMENTS

Based on the measurements carried out using the AOSCE tool, a number of observations are derived over the evolution of Java and AspectJ versions of *UAS*. The observations are dealt in detail by initially looking at the metrics and later on the higher level quality attribute, maintainability.

### A. Measurement for UAS J Version 1.0 … 1.4

Even though five metrics have been proposed to measure the symptoms of complexity, for the Java versions, only two metrics have relevance to the artifacts modeled in the case study. The first metric namely *WMCA* increases during the evolution because of increase in the number of unique functionalities modeled through methods of classes. Similarly the number of attributes in the classes also increases in the Java versions and the cross-cutting functionalities, namely, logging, persistence, exception handling, etc are scattered and tangled in the classes which is reflected by higher values of *WMCA* when compared with the equivalent AspectJ version. The value of *NAC* also increases in the Java versions during the evolution, because of adding more functionality in the later versions.

TABLE IV. UAS JAVA VERSIONS METRIC VALUES

| Version | WMCA | NAC | WPA | WAA | WJP |
|---|---|---|---|---|---|
| UAS J1.0 | 14 | 9.583 | | | |
| UAS J1.1 | 20 | 9.461 | | | |
| UAS J1.2 | 31 | 10.952 | NA | NA | NA |
| UAS J1.3 | 35 | 11.000 | | | |
| UAS J1.4 | 44 | 10.617 | | | |

### B. Measurement for UAS AJ Version 1.0 … 1.4

The AspectJ version models the cross-cutting concerns in separate units of modularity. In *UAS AJ 1.0*, the values of *WPA*, *WAA* and *WJP* are zero because refactoring of cross-cutting functionalities has not been done in this version. The values of *WPA*, *WJP*, *WAA* and *NAC* are increased in *UAS AJ* 1.1, but the values of *WMCA* decrease because the logging and persistence cross-cutting functionalities are modeled as aspects. All the metric values are increased in *UAS AJ* 1.2 because in this version, security is modeled as an aspect. The *WPA*, *WAA* and *WJP* values in version *UAS AJ* 1.4 are also increased when compared to the previous version because, we have refactored five cross-cutting concerns in this version. However, the value of *WMCA* is decreased when compared to the equivalent Java version which results in the decrease in class complexity.

TABLE V. UAS ASPECTJ VERSIONS METRIC VALUES

| Version | WMCA | NAC | WPA | WAA | WJP |
|---|---|---|---|---|---|
| UAS AJ 1.0 | 12 | 9.583 | NA | NA | NA |
| UAS AJ 1.1 | 11 | 16.909 | 1.5 | 0.8 | 0.5 |
| UAS AJ 1.2 | 21 | 15.291 | 2.4 | 1.6 | 0.8 |
| UAS AJ 1.3 | 28 | 15.300 | 3.3 | 2.0 | 1.1 |
| UAS AJ 1.4 | 32 | 13.324 | 3.9 | 2.2 | 1.3 |

### C. Aspectization Imapct on Complexity

Since, the complexity design property cannot be measure directly, metrics are used to measure its occurrences. In this work, the complexity of *AO* software was measured by using a set of metrics which affects the artifacts. The proposed metric values were found to be lesser in AspectJ versions compared to the equivalent Java versions. For example in *UAS J* 1.1 since, the functionalities contain the tangled and scattered functions, the values of *WMCA* and *NAC* are high in comparison to that of *UAS AJ* 1.1. In *UAS AJ* 1.1, the logging and persistence non-functional requirements are modeled as aspects which results in the decrease of complexity.

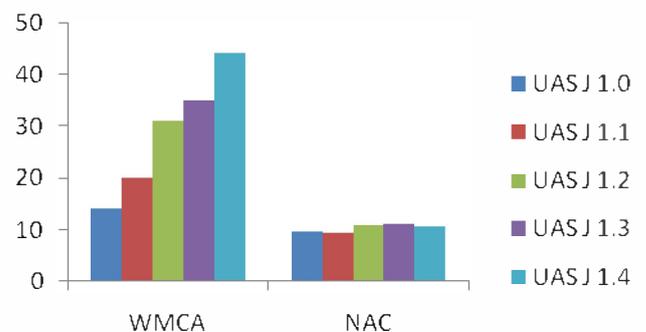

Fig. 4. Values of Proposed Metrics for UAS Java Versions

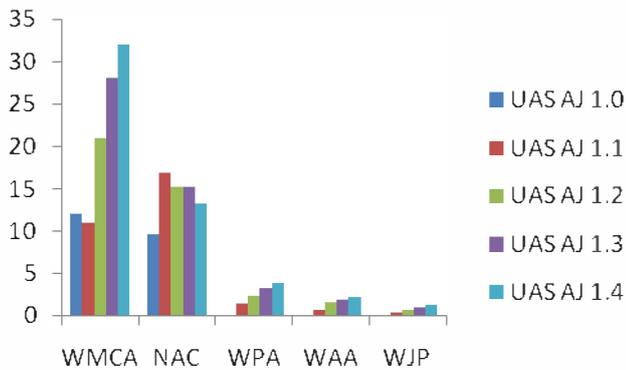

Fig. 5. Values of Proposed Metrics for UAS AspectJ Versions

*D. Inference on Maintainability*

The maintainability is an important quality attribute that captures the effort required for the smooth functioning of a software. Complexity has a direct relationship towards the maintenance of software. If a software is hard to understand and verify, it will lead to lower understandability and modifiability. Based upon the measurement, the complexity of AspectJ version of *UAS* is lesser compared to the Java version. Hence, *AO* version of *UAS* is better understandable and modifiable compared to Java version of *UAS*. Based upon this, argument, it can be inferred that *UAS AJ 1.x* versions show good characteristics of maintainability.

IX. CONCLUSIONS AND FUTURE DIRECTIONS

The primary contributor to the overhead of software expenditure is the expensive process of maintaining an existing software. Software complexity is an important design property which directly relates to its maintainability. In this paper, an *AO* complexity evaluation model has been designed to evaluate the existence of the symptoms of complexity. The case study, *UAS* is developed in both Java and AspectJ programming languages. A tool has been designed and developed to automatically measure the values of the proposed metrics. The effect of aspectization is analyzed to understand the impact on complexity. The impact on complexity across the versions of *UAS* is used to understand its influence on the maintainability of both Java and AspectJ implementations. Based on the measurements, the values of the proposed metrics in AspectJ versions were lower compared to its equivalent Java versions, and hence it can be inferred that the complexity is lower in *AO* versions of the case study. This work can be extended by identifying the most crucial artifact that can affect the complexity. The proposed model can also be applied to more cases studies with higher number of versions to generalize the impact of *AOSD* on maintainability of software.